\documentclass[conference]{IEEEtran}
\IEEEoverridecommandlockouts
\usepackage{cite}
\usepackage{amsmath,amssymb,amsfonts}
\usepackage{algorithmic}
\usepackage{graphicx}
\usepackage{textcomp}
\usepackage{xcolor}
\usepackage{hyperref}       
\usepackage{url}            
\usepackage{caption}
\usepackage{subcaption}
\def\BibTeX{{\rm B\kern-.05em{\sc i\kern-.025em b}\kern-.08em
    T\kern-.1667em\lower.7ex\hbox{E}\kern-.125emX}}
\begin{document}

\title{Towards Optimising EEG Decoding using Post-hoc Explanations and Domain Knowledge\\

}

\author{\IEEEauthorblockN{Param Rajpura}
\IEEEauthorblockA{Human-AI Interaction (HAIx) Lab \\
\textit{IIT Gandhinagar, India}\\
}

\and
\IEEEauthorblockN{Yogesh Kumar Meena}
\IEEEauthorblockA{Human-AI Interaction (HAIx) Lab \\
\textit{IIT Gandhinagar, India}}
}

\maketitle

\begin{abstract}
Decoding Electoencephalography (EEG) during motor imagery is pivotal for the Brain-Computer Interface (BCI) system, influencing its overall performance significantly. As end-to-end data-driven learning methods advance, the challenge lies in balancing model complexity with the need for human interpretability and trust. Despite strides in EEG-based BCIs, challenges like artefacts and low signal-to-noise ratio emphasise the ongoing importance of model transparency. This work proposes using post-hoc explanations to interpret model outcomes and validate them against domain knowledge. Leveraging the GradCAM post-hoc explanation technique on the EEG motor movement/imagery dataset, this work demonstrates that relying solely on accuracy metrics may be inadequate to ensure BCI performance and acceptability. A model trained using all EEG channels of the dataset achieves 72.60\% accuracy, while a model trained with motor-imagery/movement-relevant channel data has a statistically insignificant decrease of 1.75\%. However, the relevant features for both are very different based on neurophysiological facts. This work demonstrates that integrating domain-specific knowledge with Explainable AI (XAI) techniques emerges as a promising paradigm for validating the neurophysiological basis of model outcomes in BCIs. Our results reveal the significance of neurophysiological validation in evaluating BCI performance, highlighting the potential risks of exclusively relying on performance metrics when selecting models for dependable and transparent BCIs. 
\end{abstract}


\section{Introduction}

Brain-computer interface (BCI) transfers information between the brain and a computing system, encompassing the transmission of neural signals from the brain to the machine and the subsequent analysis of these cerebral responses \cite{wolpaw2002brain,hill2014general}. With a promise for their widespread adoption, portability, and cost-effectiveness, electroencephalography (EEG)-based BCIs have catalyzed applications ranging from post-stroke motor rehabilitation to the control of wheelchair systems and immersive experiences in video gaming \cite{moore2010applications,meena2015towards}. 
At the core of such applications is the analysis of EEG signals that enables the identification of patterns within EEG datasets in frequency, time, and spatial dimension. 
To identify user-invariant patterns, complex models for data-driven learning are proposed \cite{roy2019deep,craik2019deep}. However, higher performance often comes at the cost of interpretability, leading to a trust gap, especially in high-stakes applications \cite{tjoa2020survey,tonekaboni2019clinicians,ribeiro2016should}. The concept of eXplainable Artificial Intelligence (XAI) emerged to develop techniques capable of explaining model predictions. This goal seeks to facilitate human comprehension and trust in artificial intelligence (AI) systems, all while maintaining the accuracy of the models \cite{gunning2017explainable}.
While XAI has been applied in BCIs for interpretations of features learned by proposed models \cite{rajpura2023explainable,kim2023designing}, comparison of model performance is based solely on the accuracy of correctly classifying the intended activity from the EEG data. Moreover, EEG data has various artefacts and low signal-to-noise ratio (SNR) \cite{lai2018artifacts} that further necessitates model transparency to validate the prominent features learned from a limited dataset. 

Cross-participant variability presents a distinctive challenge in BCIs that significantly impacts the robustness and generalisability of these systems \cite{saha2020intra}. As a result, models trained on data from one participant may not seamlessly translate to others. Acknowledging this fact, we investigate the neurophysiological basis of model predictions to evaluate better performance and generalization of a model on a given dataset. In this work, we address model interpretability for BCIs with the following key contributions: 1) Propose a novel approach for combining domain knowledge and XAI techniques for BCIs to visualise and validate spatial, temporal and spectral domain explanations. 2) Exemplify the necessity of validating the predicted outcomes of complex models used in BCIs with neurophysiological explanations by demonstrating a use case for motor imagery and execution task classification. The remainder of this paper proceeds as follows:
\autoref{sec:Method} discusses the dataset and methods, including the architecture of the predictive model and the XAI technique. \autoref{sec:Results}  presents the results. Finally, \autoref{sec:Discussion} discusses the findings and the significance of the work.

\section{Material and Method}
\label{sec:Method}

\begin{figure}[t!]
    \includegraphics[width=0.45\textwidth]{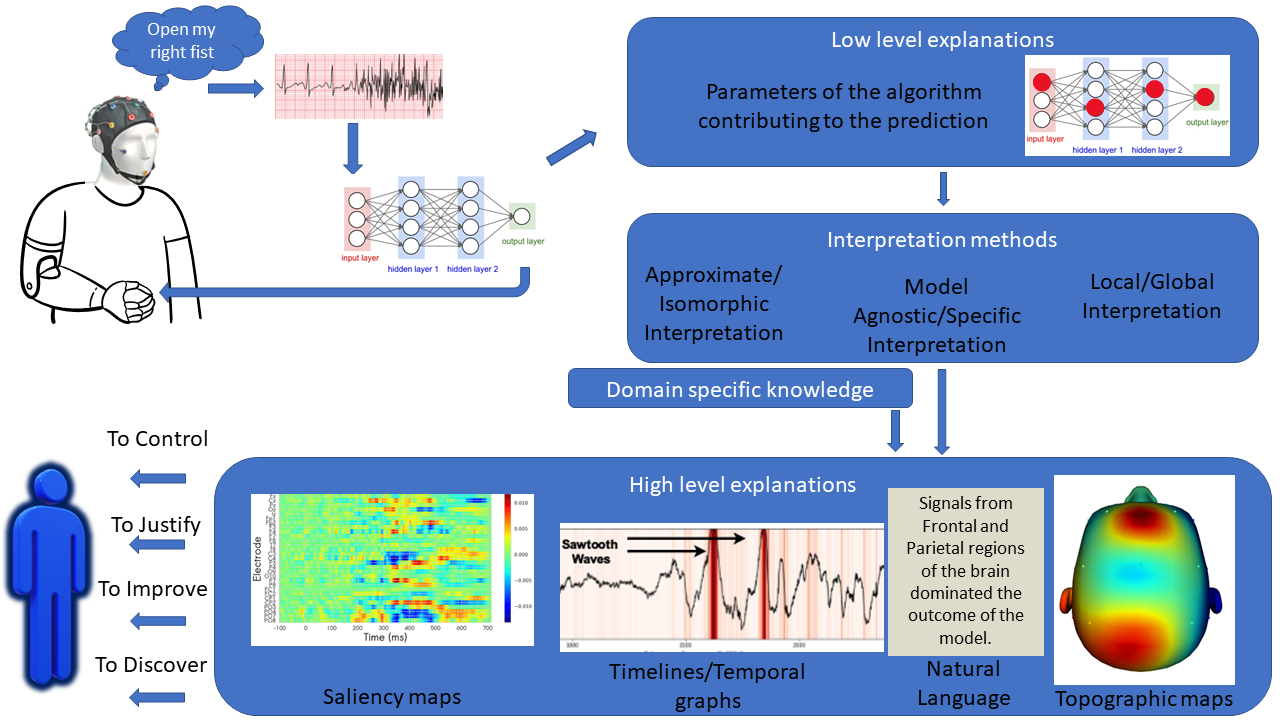}
    \caption{Overview of design space for XAI applied to Motor Imagery and Execution task. }
    \label{fig:ExampleProsthetic}
\end{figure}

\autoref{fig:ExampleProsthetic} depicts the design space for BCI assisting motor movements with prosthetics to provide an overview of this work. The design space is extended from the XAI4BCI design space proposed by Rajpura et al. \cite{rajpura2023explainable}. The dataset defines the "what" variable in the design space. The choice of predictive models and explanation framework with interfaces determines the "how" aspect. Designing the user interface, including explanation interfaces, dataset information, and model details, considers the "who" and "why" aspects of the design space. We assume that the current explanation framework, based on statistical observations, is more relevant for researchers and developers to enhance model performance and justify predictions.

\subsection{Dataset Description}
We use EEG motor movement/imagery dataset (EEGMMID-Physionet)\cite{goldberger2000physiobank,schalk2004bci2000}, one of the largest datasets on motor imagery (MI) and movement, with 64-channel EEG recordings collected from 109 participants.  We considered the two tasks: 1) open and close left or right fist and 2) imagine opening and closing left or right fist. 

We merged both tasks to augment the dataset. Each subject, on average, did 93 trials, each 4 seconds long. We divided each 4-second trial into four 1-second length runs, resulting in 160 sample recordings in each sequence and an average of 370 sequences for each subject. A bandpass filter of 8-30~Hz was applied to remove any noise in the EEG signals \cite{o2014exploring,meena2015simultaneous}. Each sequence was labeled based on the activity for left and right fist movement/imagery respectively.

\subsection{Experimental Setup}

We follow the experimental setup and implementation of EEG conformer model architecture\cite{song2022eeg} for the empirical study. The model architecture has three stages: the convolutional stage consisting of two one-dimensional convolutions to learn temporal and spatial dependencies, followed by the self-attention module to learn global dependencies, and finally, the classifier stage, two fully connected layers along with the softmax function to predict the activity in hands or limb movement/imagery. The model is trained separately for each participant with the same hyperparameters.

For the visualization and explanation interface, leveraging the statistical data-driven observations, we use Gradient-weighted Class Activation Mapping (Grad-CAM) \cite{selvaraju2017grad} to generate the feature relevance maps (montage images highlighting significant EEG channels based on feature relevance scores from GradCAM) and time-frequency charts with topography maps using Morlet wavelets\cite{tallon1997oscillatory}.

The model is trained using all 64-channel data available in the dataset. We chose 16 participants with an overall accuracy of at least 10\% higher than the chance accuracy level for reliable outcomes in the next steps. The criteria ensure that the explanations are drawn from models having significant performance on the dataset.  Further, to evaluate the effectiveness of GradCAM technique, the top 10 channels most relevant as per GradCAM scores for each left and right-hand movement/imagery are used to train the model. The reason these channels are chosen class-wise is to avoid any bias towards a single class. Finally, to compare the performance of GradCAM relevant channels with domain knowledge, the model is trained considering the 21 channels positioned near the motor cortical regions, especially central, frontal-central, and parietal regions \cite{mcfarland2000mu}.

\begin{table*}
\centering
\resizebox{\textwidth}{!}{
\begin{tabular}{c | c | c c c | c c c | c c c}
\textbf{ID} & \multicolumn{1}{c|}{\textbf{\begin{tabular}[c]{@{}c@{}}Chance level\\ Accuracy\end{tabular}}} & \multicolumn{3}{c|}{\textbf{Accuracy using all EEG channels}} & \multicolumn{3}{c|}{\textbf{\begin{tabular}[c]{@{}c@{}}Accuracy using top 17  relevant \\ channels from GradCAM\end{tabular} }} & \multicolumn{3}{c}{\textbf{\begin{tabular}[c]{@{}c@{}}Accuracy using 21 \\ MI relevant channels\end{tabular} }} \\
\hline
 &  & \multicolumn{1}{c}{\textbf{\begin{tabular}[c]{@{}c@{}}Overall\end{tabular}}} & \multicolumn{1}{c}{\textbf{\begin{tabular}[c]{@{}c@{}}Left fist\end{tabular}}} & \multicolumn{1}{c}{\textbf{\begin{tabular}[c]{@{}c@{}}Right fist\end{tabular}}} & \multicolumn{1}{|c}{\textbf{\begin{tabular}[c]{@{}c@{}}Overall\end{tabular}}} & \multicolumn{1}{c}{\textbf{\begin{tabular}[c]{@{}c@{}}Left fist\end{tabular}}} & \multicolumn{1}{c|}{\textbf{\begin{tabular}[c]{@{}c@{}}Right fist\end{tabular}}} & \multicolumn{1}{c}{\textbf{\begin{tabular}[c]{@{}c@{}}Overall\end{tabular}}} & \multicolumn{1}{c}{\textbf{\begin{tabular}[c]{@{}c@{}}Left fist\end{tabular}}} & \multicolumn{1}{c}{\textbf{\begin{tabular}[c]{@{}c@{}}Right fist\end{tabular}}} \\
7 & 59.95 & 70.97 & 69.64 & 72.97 & 65.59 & 67.86 & 62.16 & 77.42 & 71.43 & 86.49 \\
15 & 58.81 & 68.82 & 70.91 & 65.79 & 62.37 & 69.09 & 52.63 & 65.59 & 65.45 & 65.79 \\
29 & 58.29 & 69.15 & 74.55 & 61.54 & 58.51 & 76.36 & 33.33 & 70.21 & 78.18 & 58.97 \\
32 & 56.72 & 69.89 & 73.58 & 65.00 & 62.37 & 66.04 & 57.50 & 70.97 & 75.47 & 65.00 \\
35 & 55.38 & 68.82 & 65.38 & 73.17 & 66.67 & 67.31 & 65.85 & 73.12 & 80.77 & 63.41 \\
42 & 56.91 & \textbf{82.80} & 86.79 & 77.50 & \textbf{63.44} & 79.25 & 42.50 & \textbf{80.65} & 81.13 & 80.00 \\
43 & 57.72 & 69.89 & 77.78 & 58.97 & 65.59 & 68.52 & 61.54 & 63.44 & 77.78 & 43.59 \\
46 & 58.06 & 73.12 & 75.93 & 69.23 & 74.19 & 96.30 & 43.59 & 69.89 & 72.22 & 66.67 \\
48 & 56.64 & 73.12 & 81.13 & 62.50 & 66.67 & 84.91 & 42.50 & 68.82 & 67.92 & 70.00 \\
49 & 58.54 & 70.97 & 87.04 & 48.72 & 69.89 & 77.78 & 58.97 & 70.97 & 77.78 & 61.54 \\
54 & 57.99 & 70.97 & 72.22 & 69.23 & 74.19 & 83.33 & 61.54 & 78.49 & 83.33 & 71.79 \\
56 & 57.45 & 77.42 & 90.57 & 60.00 & 79.57 & 84.91 & 72.50 & 68.82 & 71.70 & 65.00 \\
62 & 58.81 & 68.82 & 74.55 & 60.53 & 58.06 & 69.09 & 42.11 & 69.89 & 76.36 & 60.53 \\
93 & 59.62 & 72.04 & 83.64 & 55.26 & 65.59 & 63.64 & 68.42 & 60.22 & 65.45 & 52.63 \\
94 & 58.27 & 82.80 & 88.89 & 74.36 & 65.59 & 72.22 & 56.41 & 75.27 & 77.78 & 71.79 \\
108 & 54.74 & 72.04 & 76.47 & 66.67 & 67.74 & 72.55 & 61.90 & 69.89 & 70.59 & 69.05 \\
\hline
\textbf{Mean$\pm$SD} & 57.74$\pm$1.41 & \textbf{72.60$\pm$4.54} & 78.07$\pm$7.48 & 65.09$\pm$7.63 & \textbf{66.63$\pm$5.68} & 74.95$\pm$8.91 & 55.22$\pm$11.24 & \textbf{70.85$\pm$5.35} & 74.58$\pm$5.51 & 65.77$\pm$9.99 \\
\hline
\end{tabular}
}
\caption{Chance level accuracy compared to the performance of EEG conformer model using 1) all 64 EEG channels, 2) using 17 most relevant EEG channels (effectively top 10 channels for each left and right fist prediction with 3 common channels) identified by GradCAM, and 3) using 21 Motor Imagery and movement-related EEG channels leveraging the domain knowledge by choosing channels near the motor cortical regions, especially central, frontal-central, and parietal regions.}
\label{tab:accTable}
\end{table*}

\subsection{Grad-CAM}
GradCAM is a class discriminative approach to interpret and compare the features learned by the model for classification by combining the gradients and the input features~\cite{selvaraju2017grad}. GradCAM-based approaches are employed for the interpretability of EEG signals for models built upon convolutional neural networks\cite{sujatha2023empirical}.

Considering that the model predicts a score $Y^c$ for each class label $c$  
i.e., right-hand and left-hand movements for a prediction on input EEG signal $X$. Then, the predicted score by the model could be formulated as: 
\begin{equation}\label{eqn:Yscore}
     Y^c = \sum_{k} w_k^c \frac{1}{Z} \sum_{i}\sum_{j} A_{ij}^k \tag{1}
\end{equation}

where, $w_k^c$ denotes the feature weights for class $c$. $A_{ij}^k$ refers to the activation of the penultimate convolutional layer at location $(i, j)$ of the feature map $A^k$ with $k$ feature maps. Since the two dimensions for raw EEG input are time and channels and convolutions in the first two layers of the EEG Conformer are one dimensional, i.e. temporal and spatial, $i$ and $j$ denote temporal and spatial dimensions, respectively.  ${Z}$ is the normalizing constant equal to the number of elements in the feature map. On partial derivation, the gradient information for a single element at location $(i, j)$ for feature map $k$ could be written as: 

$\frac{\partial{Y^c}}{\partial{A_{ij}^k}} =   \frac{w_k^c}{Z}$.

Further, summing up the gradients across all the locations $(i, j)$, 

\begin{equation}\label{eqn:sumgrad}
     \sum_{i}\sum_{j}\frac{\partial{Y^c}}{\partial{A_{ij}^k}} =   \sum_{i}\sum_{j}\frac{w_k^c}{Z} \tag{3}
\end{equation}

Since ${w_k^c}$ and ${Z}$ are independent of ${i,j}$, we can simplify it as ${w_k^c} = \sum_{i}\sum_{j}\frac{\partial{Y^c}}{\partial{A_{ij}^k}}$, since  $\sum_{i}\sum_{j}1 = Z$. These weights ${w_k^c}$ derived from the gradient information, denote the relevance of each element $A_{ij}^k$ in the activation of a given layer $k$. When combined with forward activations at each layer with $k$ feature maps, a localization map $L$ using GradCAM, highlighting the important features for a given class, can be represented as: 

\begin{equation}\label{eqn:locmap}
    {L_{GradCAM}^c} = ReLU\sum_{k}{w_k^c}{A^k} \tag{5}
\end{equation}

Later, the fine localization map is obtained with guided backpropagation \cite{springenberg2014striving} using the bilinearly interpolated coarse heatmap derived as above. 

In this work, the localization map, when visualized for spatial information in the input data, provides feature relevance for the EEG channels. Similarly, visualizing the EEG data with GradCAM over the temporal dimension, i.e. using the top relevance scores to highlight the timeframes the model considered important for prediction, gives a glimpse of the top relevant temporal features. Leveraging the domain knowledge and the importance of Mu-Beta band in analyzing brain activity during motor movements, when the raw EEG signals are converted to time-frequency plots with topography maps using Morlet wavelets \cite{tallon1997oscillatory}, the relevant timeframes based on GradCAM localisation map can be visualised with the frequency-specific intensity. Such a visualisation provides insight into the frequencies that were indirectly considered to be a relevant feature in the model prediction.


\section{Results}
\label{sec:Results}
\autoref{tab:accTable} presents the accuracy for three different scenarios, starting with the case when the model for each participant is trained using all 64 channel data. To evaluate the significance of feature relevance from GradCAM, only the top 10 EEG channels most significant for predicting left and right movement each (effectively a total of 17 channels since 3 channels are common ) are used to train the models. 
Mean accuracy across the participants is decreased significantly by $5.97\%$ (\textit{p=0.002} using Wilcoxon signed-rank test). 

To compare the neurophysiological relevance and leverage the domain knowledge, channels closer to brain regions involved in motor movements are considered to train the models, and their performance is reported in \autoref{tab:accTable}. The performance is decreased by $1.75\%$ but is statistically insignificant. In comparison, it is  $4.22\%$ higher than using the most relevant channels from GradCAM (data-driven approach). 

\begin{figure}[htbp!]
     \centering
     \begin{subfigure}[b]{0.24\textwidth}
         \centering
         \includegraphics[trim={15cm 0cm 12cm 2.2cm},clip,width=\textwidth]{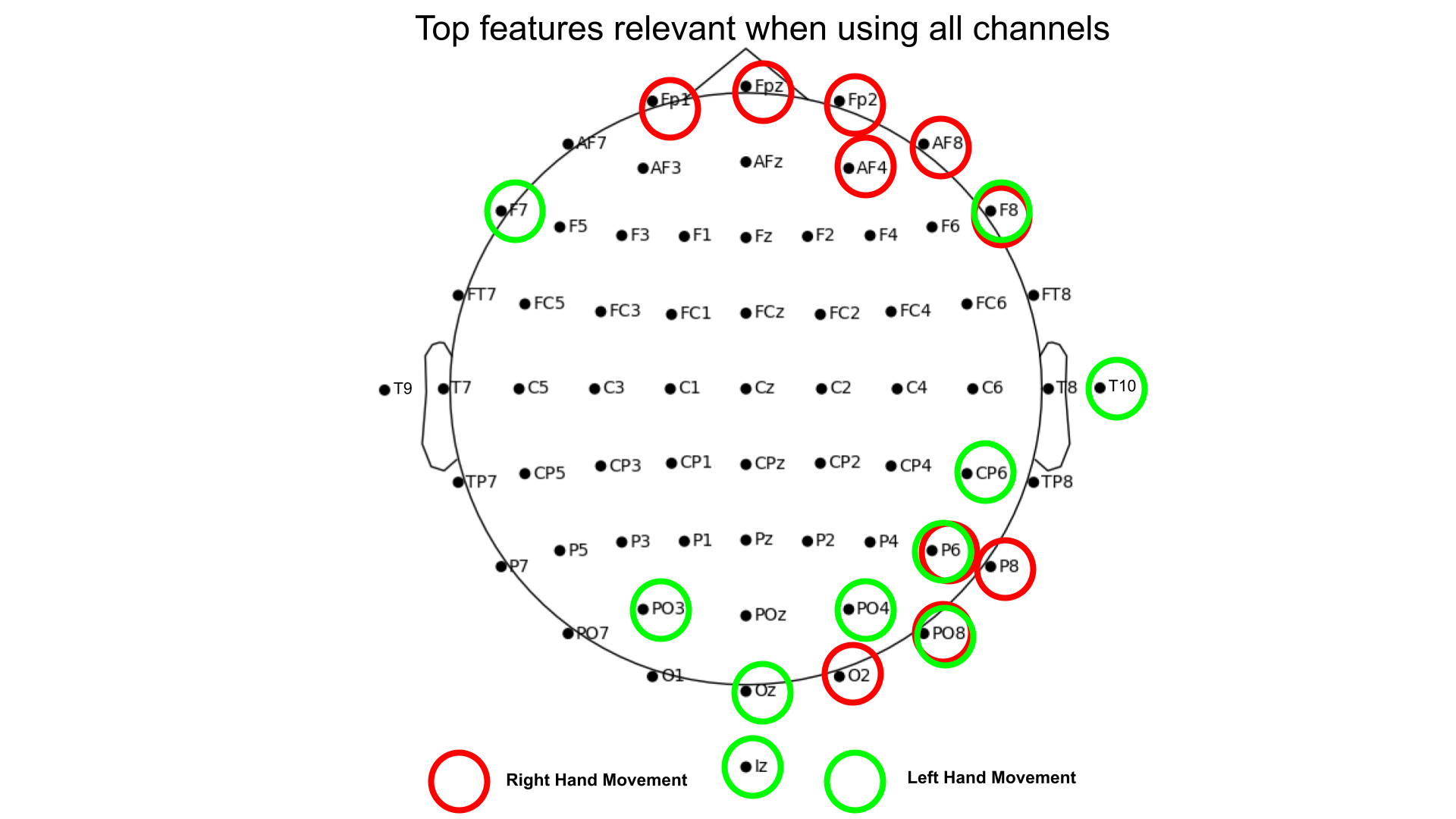}
         \label{fig:allchannelTopFR}
     \end{subfigure}
     \begin{subfigure}[b]{0.24\textwidth}
         \centering
         \includegraphics[trim={15cm 0cm 12cm 2.2cm},clip,width=\textwidth]{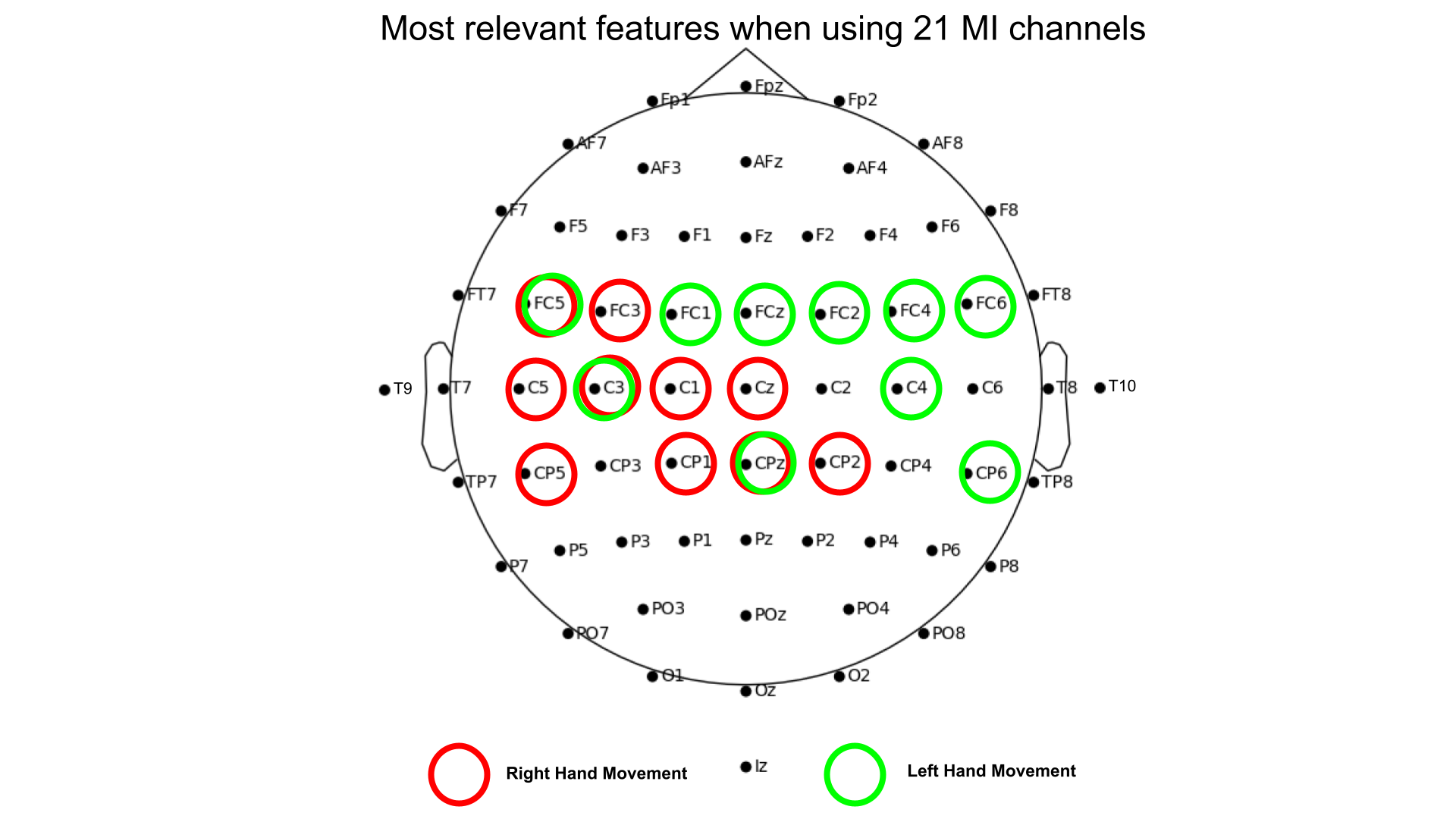}
         \label{fig:MIchanneltopFR}
     \end{subfigure}
        \caption{Montage image highlighting the significant channels identified by GradCAM when model is trained with all 64 EEG channels (Left) and 21 EEG channels (Right) relevant for Motor Imagery and movement.}
        
        \label{fig:spatial}
\end{figure}

\autoref{fig:spatial} shows the comparison of the most significant channels for predictions using GradCAM when models are trained with all EEG channels vs. relevant channels for motor movements via the EEG montage images. These significant channels are identified by calculating the frequency of an EEG channel appearing in the top 10 significant channels for right and left-hand movement predictions, respectively, across the selected 16 participants' data.

\begin{figure}[htbp!]
     \begin{subfigure}[b]{0.45\textwidth}
         \centering
         \includegraphics[trim={2cm 31cm 1cm 26cm},clip,width=\textwidth]{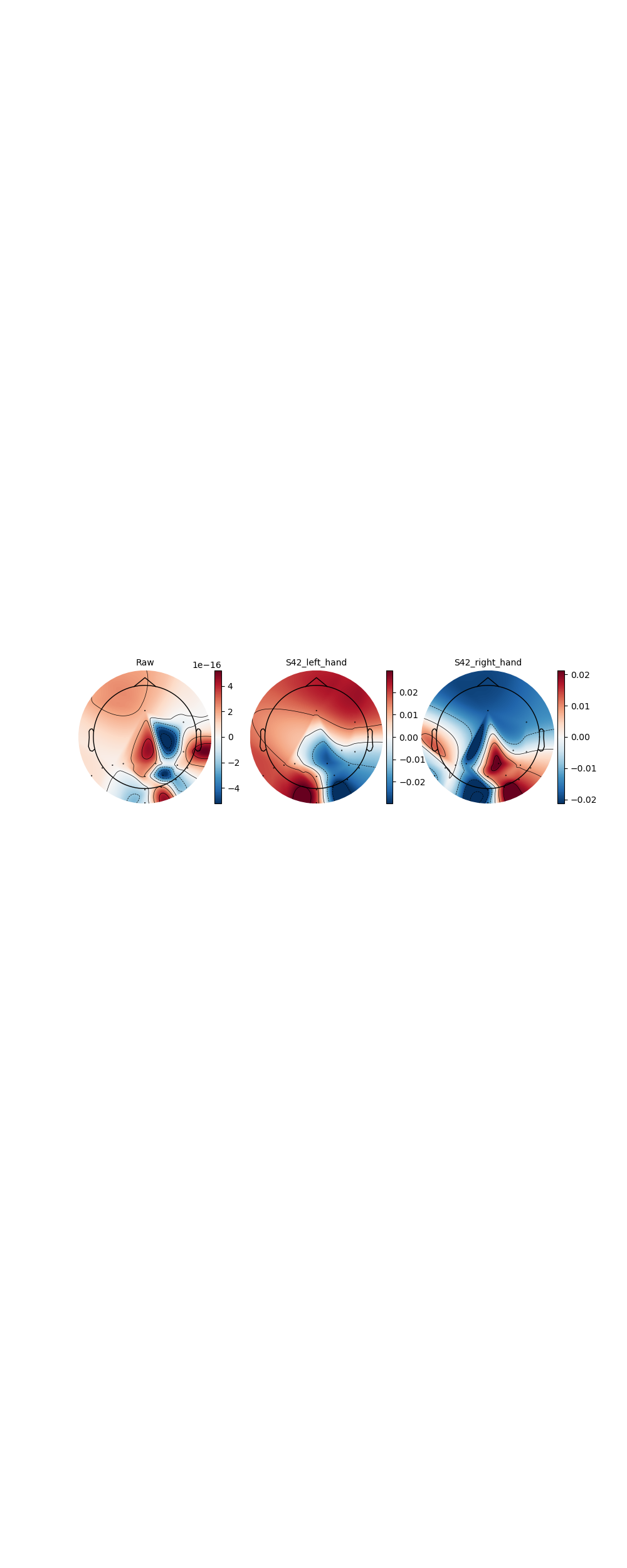}
         \label{fig:topoTopFR}
     \end{subfigure}
     \begin{subfigure}[b]{0.45\textwidth}
         \centering
         \includegraphics[trim={2cm 31cm 1cm 26cm},clip,width=\textwidth]{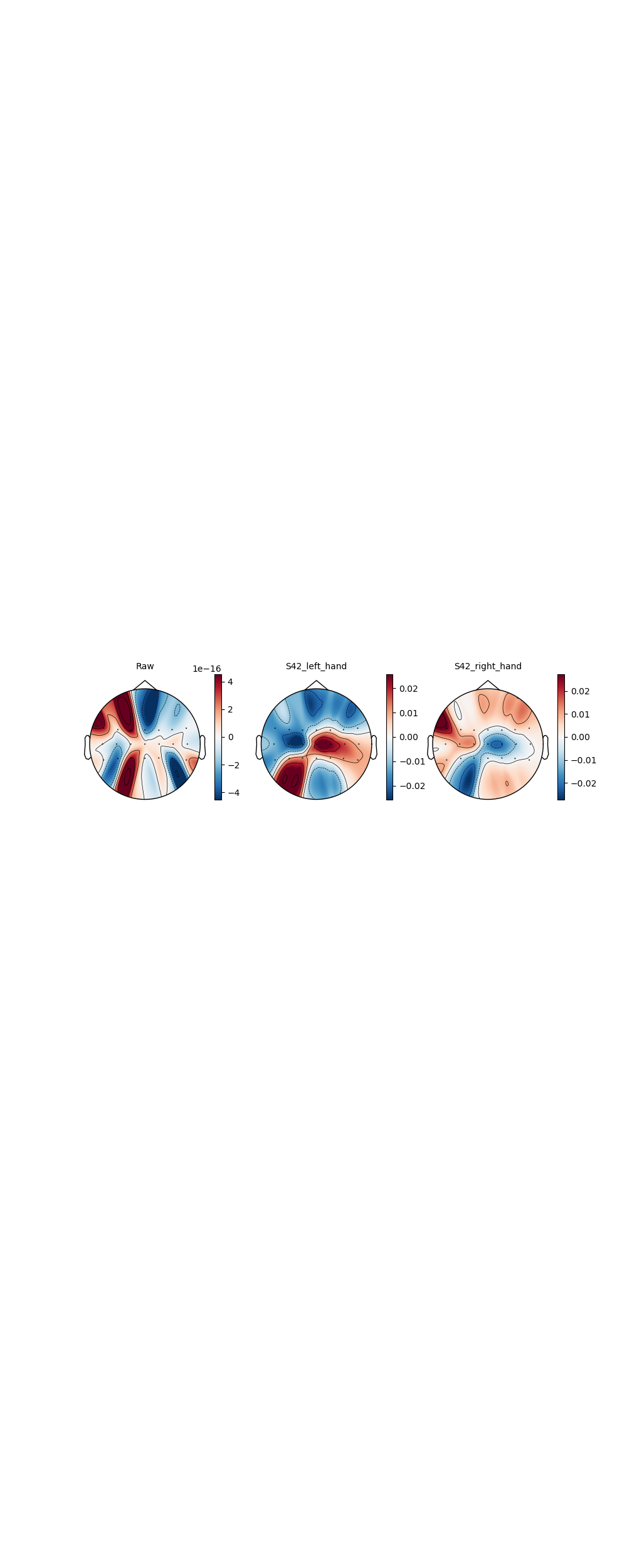}
         \label{fig:topoMI}
     \end{subfigure}
        \caption{Raw activations (Left column) during the task compared with Class Activation Topography(CAT) for both left(Centre column) and right (Right column) predictions for Subject ID 42 using top 17 relevant channels via GradCAM(Top row) and 21 MI-relevant channels(Bottom row).}
        \label{fig:topomaps}
\end{figure}

\begin{figure}[htbp!]
     \begin{subfigure}[b]{0.49\textwidth}
         \centering
         \includegraphics[width=\textwidth]{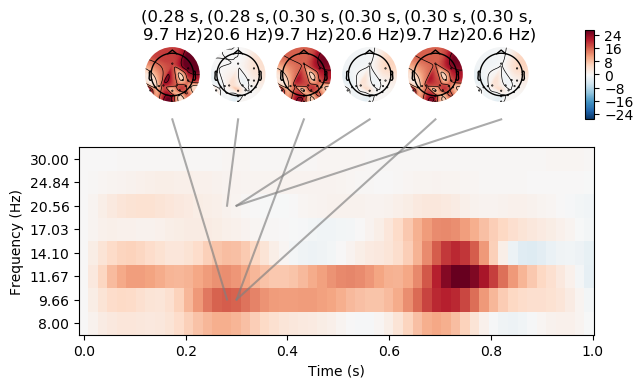}
         \label{fig:indTimeFreqTopFR}
     \end{subfigure}
     \begin{subfigure}[b]{0.49\textwidth}
         \centering
         \includegraphics[width=\textwidth]{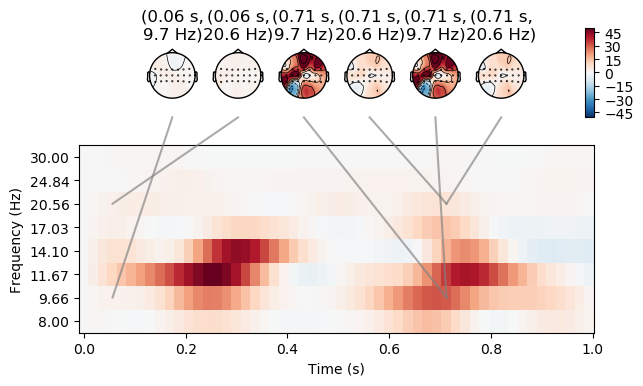}
         \label{fig:indTimeFreqMI}
     \end{subfigure}
        \caption{(Top) Time-frequency plots with topo maps in the Mu-Beta band depicting relevant frequencies during right fist movements, marked by GradCAM when trained with the top 17 channels. For Subject ID 42, model accuracy dropped to $42.50\%$ when trained with the top 17 channels. (Bottom) In contrast, the bottom plots explain the model outcomes, where the model achieved $80.00\%$ accuracy for right fist movements when trained with 21 MI-relevant channels. This comparison highlights the difference in relevant timeframes affecting model predictions.}
        \label{fig:tfplots}
\end{figure}

To validate the applicability of XAI on the participant level, we chose Subject ID 42, whose accuracy dropped significantly by $17.20\%$, especially for right fist movements (drops by $37.5\%$) when using top relevant channels from GradCAM compared to the accuracy of 21 MI-relevant channels. 
\autoref{fig:topomaps}(Top) compares the Class Activation Topography (CAT) \cite{song2022eeg} with the raw activation to showcase the feature relevance across the activity period across trials for Subject ID 42 when the model is trained using 17 most significant EEG channels identified from GradCAM feature relevance scores. \autoref{fig:topomaps}(Bottom) represents the same when the model is trained using 21 MI-relevant channel data. 
\autoref{fig:tfplots}(Left) illustrates the time-frequency plots with topo maps for the frequencies in the Mu-Beta band at timestamps considered relevant by GradCAM for predictions of right-hand movements when a model is trained using 17 most significant EEG channels identified from GradCAM feature relevance scores. \autoref{fig:tfplots}(Right) illustrates the same when a model is trained using 21 MI-relevant channels. The model trained on MI-relevant channels learned to consider the 0.1s and 0.7s time windows as significant timeframes for prediction. These results signify the importance of class discriminative explanations using GradCAM that can help understand and debug models used in BCIs.  

\section{Discussion and Conclusion}
\label{sec:Discussion}



To highlight the significance, our results can be perceived from three levels. When a sole data-driven approach is used, the performance can be represented as in \autoref{tab:accTable}. The metrics reflect that all the models, with or without XAI or domain knowledge, perform similarly. Such an analysis based on performance metrics limits understanding of what the data-driven models learn to predict accurately. However, \autoref{fig:spatial} explains the spatial domain, visually indicating the relevance of contralateral activation in predicting the target tasks when the model is trained on MI-relevant channels. At this level, validation from the existing domain knowledge is possible. Such an insight adds to the transparency and reliability of the results. Although the insight in this work is limited to the dataset and the model architecture, it can undoubtedly be validated with user studies to understand the expert's opinion on the model reliability and their preferences when there is an ambiguous prediction.  

Diving deeper into participant-level analysis, for participant S42, \autoref{fig:topomaps} and \autoref{fig:tfplots} indicates the relevance of event-related desynchronization and synchronization \cite{mcfarland2000mu} being efficiently captured in the better-performing model trained on MI-relevant channels. The detailed analysis and explanation provide a thorough peek to interpret the model outcomes. The approach also serves as a reliable tool for human-AI collaboration where human stakeholders, without over-reliance or bias, can inspect the model predictions. A relevant direction for this vision is to understand the needs of the stakeholders for explanations that augment their decision-making \cite{kim2024stakeholder}. Kim et al. \cite{kim2023designing} investigate the requirements for explanations from domain experts in BCI and design an interface catering the BCI domain.

While the practice of channel selection in EEG tasks is well-known, our motivation is to highlight the necessity of validating data-driven models that often do not use artefact removal techniques for end-to-end learning. Many studies have recently employed XAI techniques, but the goal of building transparent and trustworthy BCIs is incomplete without quantitative and qualitative benchmarks comparing explanations across model architectures and datasets. Further research by creating benchmarks for post-hoc and ante-hoc explanations across datasets, models, and different XAI techniques would test the extent of generalizability of the findings and potentially contribute to building robust and trustworthy BCIs.

\section{Acknowledgment}
\label{sec:acknowledgment}
This work was supported by Indian Institute of Technology Gandhinagar startup grant IP/IITGN/CSE/YM/2324/05. Our source code is publicly available on \href{https://github.com/HAIx-Lab/PosthocXAI4BCI}{https://github.com/HAIx-Lab/PosthocXAI4BCI}. 
\bibliography{references}

\end{document}